# Comment on "Systematic survey of high-resolution b-value imaging along Californian faults: inference on asperities"

Yavor Kamer[1, 2]

[1]Swiss Seismological Service, ETH Zürich, Switzerland

[2]Chair of Entrepreneurial Risks, Department of Management, Technology and Economics, ETH Zürich, Switzerland

[*Tormann et al.*, 2014] propose a distance exponential weighted (DEW) b-value mapping approach as an improvement to previous methods of constant radius and nearest neighborhood. To test the performance of their proposed method the authors introduce a score function:

$$score = \frac{N}{n^2} \sum_{i}^{n} abs(b_{true} - b_{est}) \qquad (1)$$

where $N$ is the total number of grid nodes, $n$ is the number of non-empty nodes, $b_{true}$ and $b_{est}$ are the true and estimated b-values. This score function is applied on a semi-synthetic earthquake catalog to make inference on the parameters of the method. In this comment we argue that the proposed methodology cannot be applied to seismic analysis since it requires a priori knowledge of the spatial b-value distribution, which it aims to reveal.

The score function given in Equation 1 seeks to minimize the absolute difference between the generating (true) and estimated b-values. Like any statistical parameter, the estimation error of the b-value decreases with increasing sample size. However, since the b-value is a measure of slope on a log-log plot, for any given sample size the estimation error is not only asymmetric but also its amplitude is dependent on the b-value itself.

To make a pedagogical analogy; conducting b-value analysis by limiting the sample size (Tormann et al. use 150) is similar to measuring temperature with a peculiarly short thermometer. For small sample sizes such a thermometer would measure lower temperatures more precisely than higher ones, to the extent that it would be impossible to distinguish between values on the upper end. Only when the sample size, and hence the resolution, is increased does the thermometer become reliable across the whole measurement range. For an illustration in Figure 1a we show the confidence intervals for 6 b-values, accordingly we divide our thermometer into 6 intervals. The length of each interval is scaled with $1/\sigma_N$ as a proxy for resolution, where $\sigma_N$ is the standard deviation for N number of samples. The analogous thermometers obtained for N=150, 300 and 1500 are shown in Figure 1b.

Intersecting the confidence interval curves of different b-values one can derive that the crudest measurement in the range [0.6-1.5] would require a resolution of $\Delta b=0.3$ that is achieved with at least ~170 samples. An analysis in this constrained setting can correspond only to a mere classification of b-values as low (0.70±0.10), medium (1.00±0.13) or high (1.30±0.17). Such a graphical inference is, however, an oversimplification because determining if two GR distributions are significantly different (for a given *p*-value) would require nested-hypothesis tests (see Supporting Document)

Compared to the previous sampling techniques the proposed DEW method features two additional parameters; number of maximum events ($N_{max}$) and coefficient of exponential magnitude weight decay with distance ($\lambda$). The remaining parameters, common with the previous approaches, are set to the following values: minimum number of events ($N_{min}$=50), maximum search radius ($R_{max}$=7.5) and minimum search radius to find closest event ($R_{min}$=2.5km). The authors propose a synthetic Monte-Carlo approach to indentify which parameter values are superior in retrieving the input structure. Their synthetic catalog features three different b-value regions (0.5, 1.3 and 1.8) embedded in a background of b=1. The input b-value of the dominant region (both in terms of number of events and non-empty nodes) is chosen as 0.5. The authors find the best value for $\lambda$ to be 0.7 while they set the $N_{max}$ to a constant value of 150. Considering the confidence intervals for different b-values (Figure 1a) such small sample sizes and large weight decay may be sufficient for small b-values, however, for larger b-values the same parameter values would be sub-optimal. In order to illustrate this we obtain the same dataset used by Tormann et. al. and conduct the same numerical test varying the input b-value of the dominant region from 0.5 to 1.25 (Figure 2). We vary the parameters $N_{max}$ and $\lambda$ in the ranges of [150-1950] and [0.01-1]. For each parameter pair the score function was averaged over 500 realizations of the synthetic catalog. The score function surfaces for the 4 different input b-values are presented in Figure 2a. Although in their synthetic test Tormann et. al. conduct their optimization procedure by setting $N_{max}$ =150, in the application to real seismicity the authors set $N_{max}=\infty$; effectively designating the limiting parameter as $R_{max}=7.5$. This parameter flip does not change our following conclusion, however, since for any spatial distribution $N_{max}$ and $R_{max}$ are coupled via its fractal dimension [*Hirabayashi et al.*, 1992]. Nevertheless we repeat the synthetic test setting $N_{max}=\infty$ and varying $R_{max}$ in the range of [1-100km]. The score surfaces are presented in Figure 2b. We observe that the minima vary as a function of the synthetic input: structures with b=1 (which is one of the most well established empirical laws in seismology [*Gutenberg and Richter*, 1954]) require much larger $N_{max}$ (i.e $R_{max}$) and lower $\lambda$ values to be retrieved correctly. This indicates that choosing parameter values based on a synthetic input with low b-values will lead to under sampling of regions with higher b-values. Felzer (2006) observed that for a uniform b-value of b=1 such under sampling leads to emergence of artifacts with high or low b-values. Similar concerns regarding the under sampling of the Gutenberg-Richter's distribution have been brought up numerous times [*Shi and Bolt*, 1982; *Frohlich and Davis*, 1993; *Kagan*, 1999, 2002, 2010; *Amorese et al.*, 2010]. We remind the reader that in this comment we are not concerned with the question whether the b-value on a certain fault is uniform or not, rather we are argue that arbitrary parameter choices and improper optimization schemes can indeed lead to under sampling.

To illustrate the sensitivity of the published results to parameter choices we consider the observed Parkfield seismicity. We apply the DEW method to obtain two b-value maps; a) using the same parameter values as Tormann et al. (2014) $R_{max}= 7.5$, $\lambda=0.7$ b) using the parameters that would have been obtained if the input synthetic b-value was chosen as b=1 rather than b=0.5; $R_{max}= 40$km, $\lambda=0.01$ (Figure 3). It is important to note that Tormann et al. (2014) use these maps to calculate annual probabilities of M6 or larger event to occur. Figure 3 suggest that the emergence of high and low b-value anomalies is a mere artifact of under sampling. These artifacts lead to differences of up to two orders of magnitude in the recurrence times thus it would be precarious to use such maps for assessment of probabilistic seismic hazard on faults. Since one cannot know in advance what b-value the real data features and thus choose parameters accordingly, we maintain that the approach presented by the Tormann et al. (2014) cannot be used on real datasets as its results depend on the assumed input b-values chosen to derive its parameters. Our conclusion applies also to the previously used similar b-value mapping methods of constant radius and nearest neighborhood.

We remind the reader that spatial/temporal b-value mapping has been previously tackled with likelihood based approaches that take into account model complexities [*Imoto*, 1987; *Ogata and Katsura*, 1993]. To put the term "model complexities" into perspective we note that for the 4174 complete events in Parkfield, Tormann et al. report ~1000 spatially varying b-values saturated in the range [0.50-1.50]. Whereas Imoto (1987), using penalized likelihood optimization for a comparable sized dataset (3035 complete events in New Zealand), obtains only 28 b-values in the range of [0.94-1.08].

**Acknowledgements**

**FIGURES**

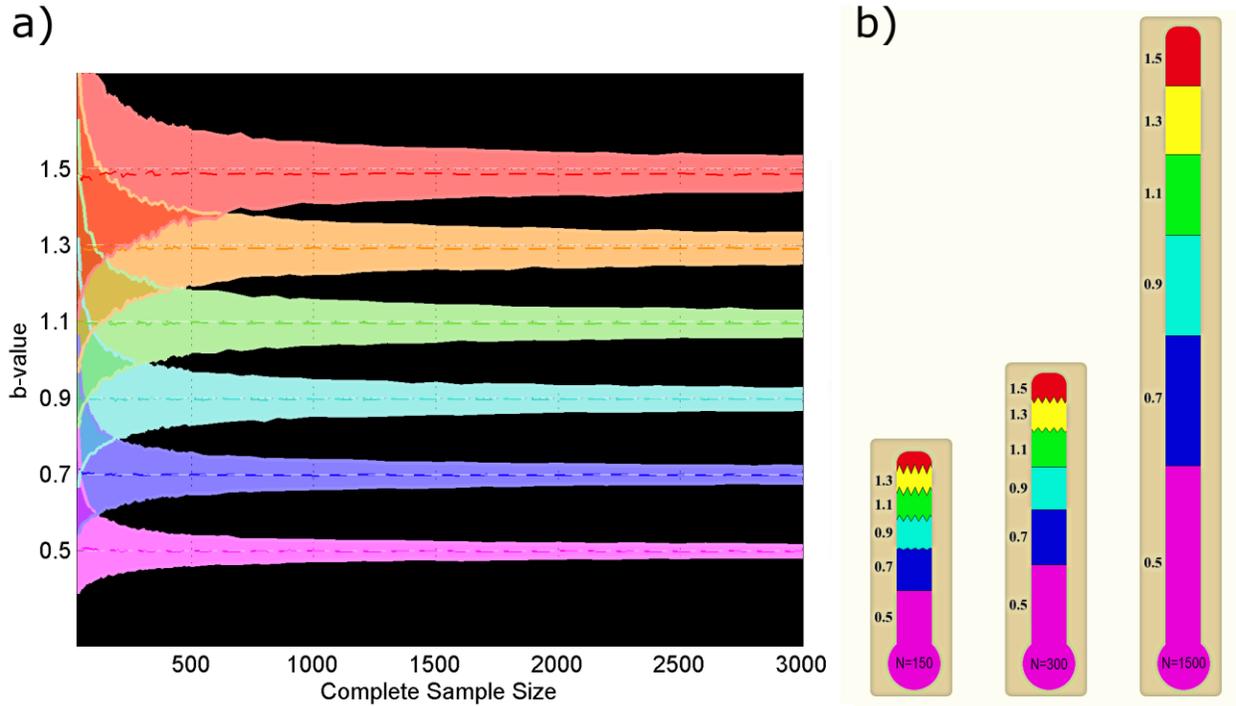

**Figure 1.** a) Confidence intervals of 5% and 95% for b-value estimations with increasing sample size. b) Analogous thermometers corresponding to sample sizes N=150, 300 and 1500. The length of each thermometer is representative of its absolute resolution. When present, confidence interval overlaps between gradations are shown by sawtooth waves with amplitudes scaled accordingly.

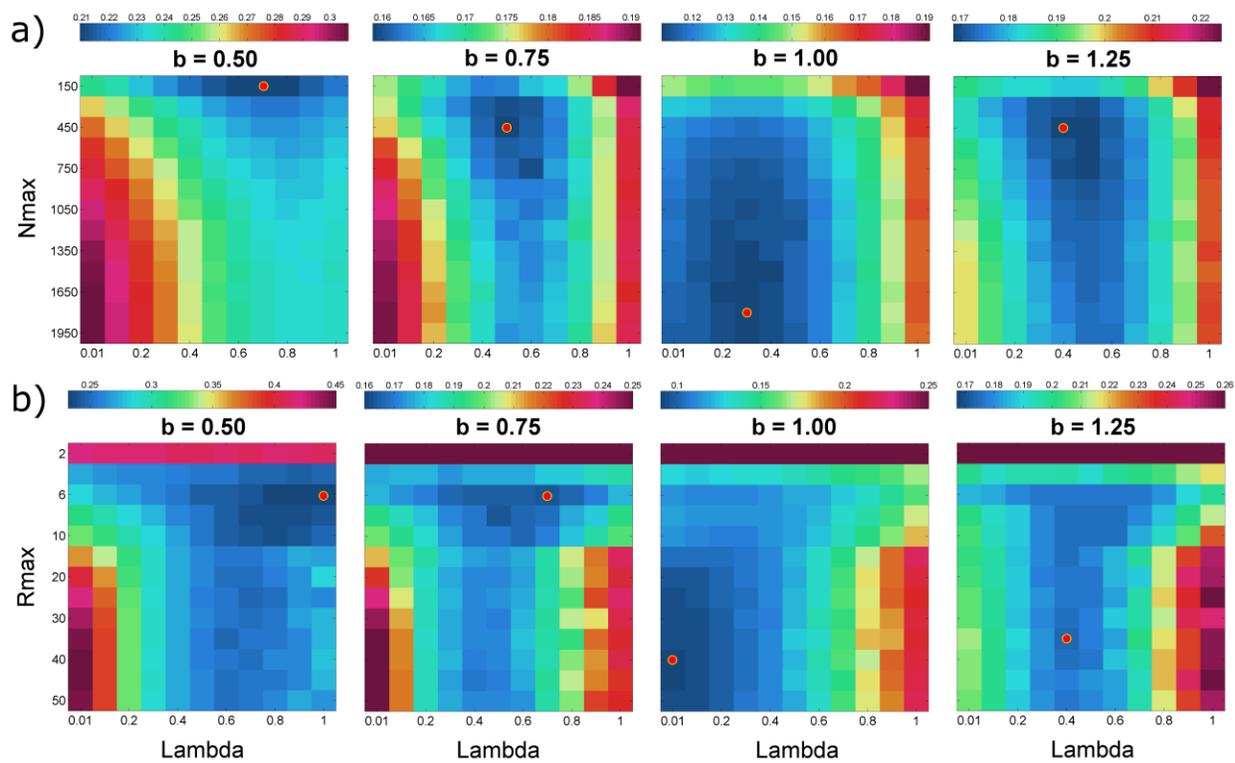

**Figure 2.** Variations of the minimum (red dots) of the score function for different input b-values as a function of a) weight decay *Lambda* and the maximum allowed sample size $N_{max}$ b) *Lambda* and the maximum search radius $R_{max}$

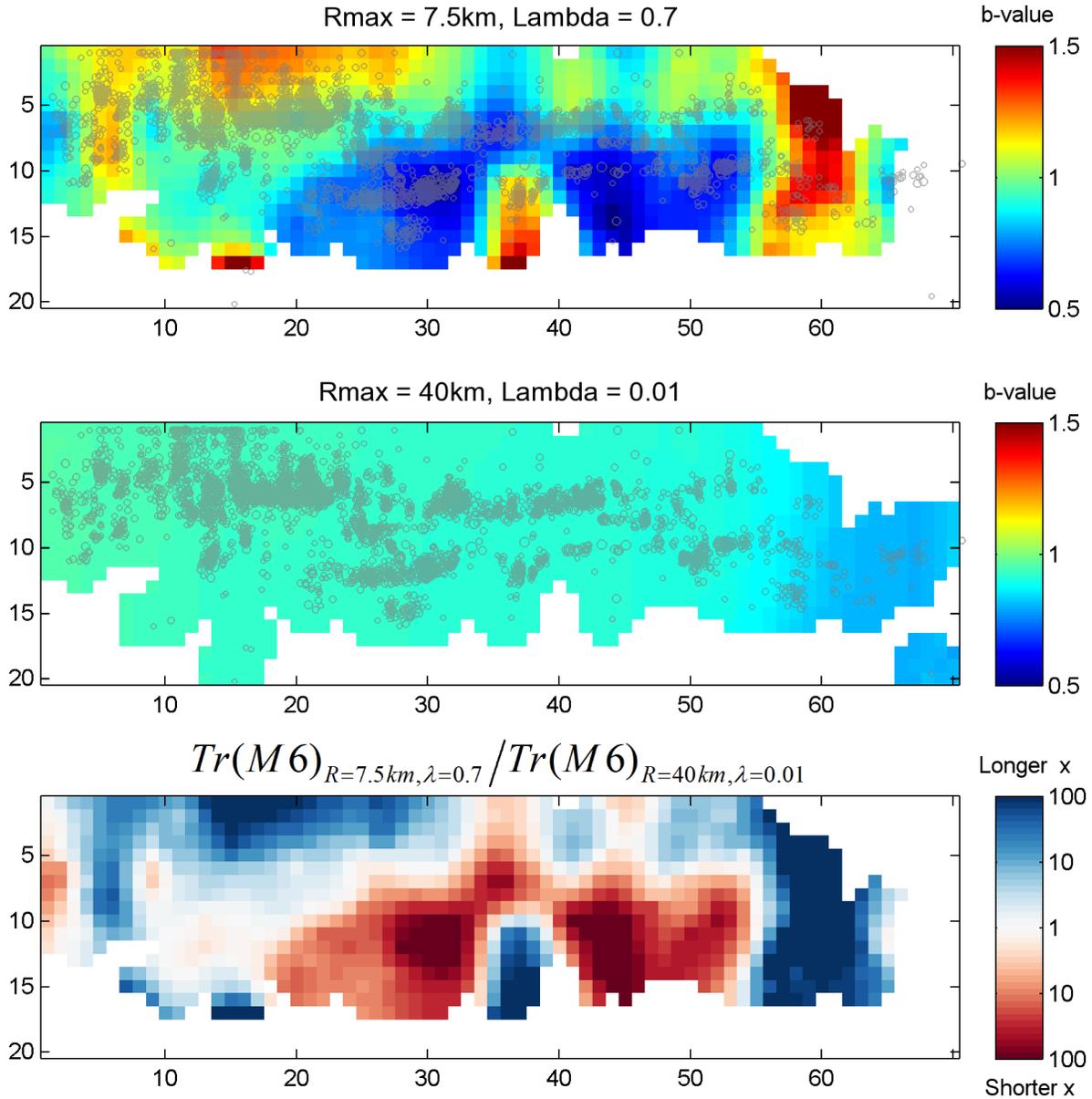

**Figure 3.** Two b-value maps obtained from the same M≥1.3 Parkfield seismicity of the last 30 years: Using the same parameters as Tormann et. al. (top panel: 0.50<b<1.86) and the parameters minimizing the score function for a synthetic input of b=1 (center panel: 0.80<b<0.96). The observed seismicity is superimposed as gray circles scaled according to magnitude. Notice that in top panel the extreme b-values are observed mainly in regions with low seismic density. Bottom panel shows the ratio of the expected recurrence times for a magnitude M6 or greater event.